\documentclass[a4paper,12pt]{article}
\usepackage{cancel}
\usepackage{ulem}
\usepackage{amsfonts}
\usepackage{amssymb}
\usepackage{graphicx}
\usepackage{amsmath}
\usepackage{enumerate}
\usepackage{mathtools}
\usepackage{subfig}
\usepackage{color}
\usepackage{float}
\usepackage{here}
\usepackage{cite}
\usepackage{mathrsfs}
\usepackage{float,epsfig}
\usepackage{dcolumn}% Align table columns on decimal point
\usepackage{tikz}
\usetikzlibrary{decorations.pathreplacing}
\usepackage{graphicx}% Include figure files
\usepackage{bm}% bold math
\usepackage{amsmath,amssymb,amsthm}
\usepackage[colorlinks=true,linkcolor=blue,citecolor=red]{hyperref}
\makeatletter \@addtoreset{equation}{section}

\newcommand{\be}{\begin{equation}}
\newcommand{\ee}{\end{equation}}
\newcommand{\bea}{\setlength\arraycolsep{2pt} \begin{eqnarray}}
\newcommand{\eea}{\end{eqnarray}}

\def\0{{\sst{(0)}}}
\def\1{{\sst{(1)}}}
\def\2{{\sst{(2)}}}
\def\3{{\sst{(3)}}}
\def\4{{\sst{(4)}}}
\def\5{{\sst{(5)}}}
\def\6{{\sst{(6)}}}
\def\7{{\sst{(7)}}}
\def\8{{\sst{(8)}}}
\def\sst#1{{\scriptscriptstyle #1}}

\begin{document}
	
\title{\bf \Large On Phase  Transition  Behaviors of  Kerr-Sen  Black Hole }
\author{ \small   A. Belhaj$^{1}$\footnote{belhajadil@fsr.ac.ma},  A. El Balali$^{1}$,  W. El Hadri$^{1}$, H. El Moumni$^{2}$\thanks{hasan.elmoumni@edu.uca.ma},
	M. A. Essebani$^{3}$,  M. B. Sedra$^{3,4}$\footnote{ Authors in alphabetical order.}
	\hspace*{-8pt} \\
	%EndAName
	{\small $^1$ D\'{e}partement de Physique, Equipe des Sciences de la mati\`ere et du rayonnement, ESMaR}\\
{\small   Facult\'e des Sciences, Universit\'e Mohammed V de Rabat, Rabat,  Morocco} \\
	{\small $^{2}$  EPTHE, D\'{e}partement de Physique, Facult\'e des Sciences,   Universit\'e Ibn Zohr, Agadir, Morocco} \\
	{\small $^{3}$  D\'{e}partement de Physique, LabSIMO,   Facult\'{e}
		des Sciences, Universit\'{e} Ibn Tofail, K\'{e}nitra,
		Morocco } \\  {\small  $^4$ Facult\'e des Sciences et Techniques d'Errachidia, Universit\'e Moulay Ismail, Errachidia, Morocco}\\
	{\small  .}\\
} \maketitle

%\date{}
%\vspace{-3.6em}
%
%\begin{center}
%\textit{$^1$
%\\ [0.5em]
%\end{center}
%
%\vspace{1em}

	\begin{abstract}
		{\noindent}
We investigate     phase transitions and critical behaviors   of the Kerr-Sen black hole in four dimensions.
 Computing the involved thermodynamical quantities including the  specific  heat and using the Ehrenfest  scheme,
   we  show that such a black hole  undergoes  a second-order phase transition. Adopting a new  metric form derived from
    the Gibss free energy scaled by a conformal factor associated with extremal solutions,
     we   calculate   the   geothermodynamical  scalar curvature
    recovering  similar   phase transitions.
     Then,  we  obtain   the  scaling laws and    the  critical  exponents,  matching perfectly with   mean field theory.
	\end{abstract}

	\newpage	
	
	\tableofcontents
	
	\newpage
	
	\section{Introduction}
It has been shown that black holes can be considered as exotic objects predicted by various  gravity theories \cite{w1,w2}.
The associated investigations  have been supported by low-energy limits of superstring models and  M-theory \cite{w3,w4,w5}.
  In this context, many solutions have been elaborated using either brane physics or the dimension reduction mechanism on non trivial geometries \cite{w6}.
A particular emphasis  has been put on a solution obtained from the  heterotic superstring theory \cite{w7}. According to some stringy assumptions,
the Kerr-Sen black hole  has been found as an exact solution   having a finite amount of charge and angular momentum.
  This model  has been approached from many angles including the thermodynamical and the optical aspects \cite{w8,w9,w10,ww}.
  Recently, such behaviors  have been considered as fascinating research topics   supported by  the black hole  imaging
   \cite{w11}. In particular,  many  studies   have  been done dealing with thermodynamics of  four dimensional  black holes
    \cite{w12,w13,w14,w15}.  Among others,   it has been shown that certain models    can undergo phase transitions which
     have been extensively investigated  in different situations and backgrounds including dark  sectors \cite{w16,w17,ww2}.
     In this way, various  thermodynamical relations  have been exploited to study the nature of such transitions.
    Precisely,  the Ehrenfest equations  have been considered  as an elegant tool to inspect the second-order  phase transition     at certain critical points \cite{w18,w19}. At the vicinity   of such   points, certain    thermodynamic quantities, including
      the  heat capacity,  exhibit singular behaviors  relaying on   divergencies.

The aim of this paper is to  investigate    phase transition behaviors  of the Kerr-Sen black hole in four dimensions by  determining
  their  nature using two different approaches.   Concretely,  we  compute   the relevant  thermodynamical quantities including the  specific
    heat involving  non trivial behaviors at the critical point. Using  the  Ehrenfest scheme,  we first   reveal   that such a black hole,  being  described by
three parameters (mass $M$, norm of angular
momentum $J$ and charge $Q$),   undergoes
      a second-order phase transition.  Adopting a new metric form  relying on the Gibbs free energy scaled by a conformal factor
       associated with extremal solutions,
       we elaborate  the geothermodynamics
         recovering similar  phase transition behaviors. Then, we obtain   scaling laws    and  critical  exponents of such a black hole,  which
         match  perfectly with  mean field theory.

The organisation of   this work is as follows. In section 2, we reconsider the study of   the phase transitions
 of  the Kerr-Sen black hole  by computing the relevant thermodynamical  quantities. In section 3,  we examine the Ehrenfest equations
   and show that the Kerr-Sen  black hole  undergoes  a second-order phase transition. In section 4,
    we compute thermodynamical curvature showing similar phase transition behaviors.  In section 5,
     we   discuss   the scaling laws and  the critical  exponents by  computing the associated quantities.
     The last section is devoted to conclusions and open questions.

\section{Kerr-Sen black hole  thermodynamics}
\begin{figure} [h]
\begin{center}
 \includegraphics[scale=0.6]{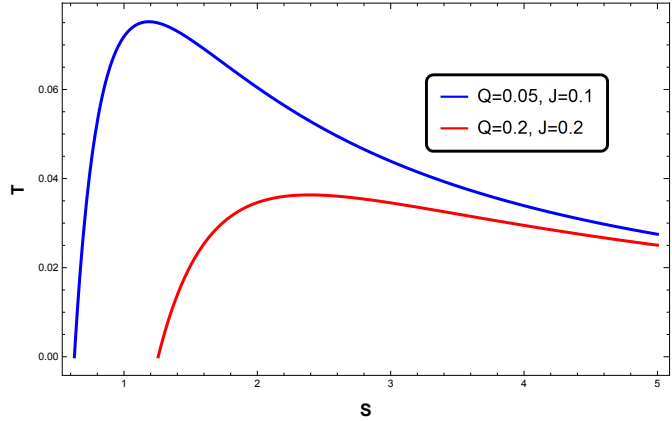}
\caption{The Temperature (T) for Kerr-Sen black hole as a function of the entropy (S) for
different values of  $J$ and $Q$.}
 \label{1}
 \end{center}
\end{figure}

\begin{figure}[h]
\begin{center}
 \includegraphics[scale=0.8]{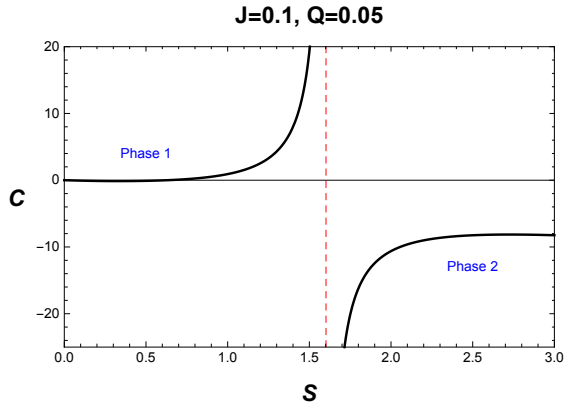}
 \caption{ The heat capacity $(C)$  of  Kerr-Sen black hole with respect to the entropy $(S)$ for
fixed $J=0.1 $ and $Q=0.05$.}
 \label{2}
 \end{center}
\end{figure}

In this section, we  reconsider  the study of the Kerr-Sen black hole thermodynamics.  This four dimensional black hole
 solution has been obtained from  the heterotic superstring  theory living in ten dimensions \cite{w7}. The associated action  is given by
\begin{equation}
S= \int d^4x\sqrt{-\det g}e^{-\Phi}\left(R- g^{\mu\nu}\partial_\mu\Phi \partial_\nu\Phi- {1\over 8}  F_{\mu\nu} F^{\mu\nu}-{1\over 12}
H_{\mu\nu\rho} H^{\mu\nu\rho}\right).
\label{S1}
\end{equation}
Here,  $R$ is the   Ricci scalar  curvature, $g$  denotes  the determinant of the metric, $F_{\mu\nu}$ indicates  the abelian electromagnetic
 tensor given by   $F_{\mu\nu} = \partial_\mu
A_\nu -\partial_\nu A_\mu$ with the Maxwell field $A_{\mu}$.  $\phi$ is the  dilaton slacar field, while
 the  tensor field  $H_{\mu\nu\rho}$  is given by
\begin{equation}
H_{\mu\nu\rho} = \partial_\mu B_{\nu\rho}+ \partial_\nu B_{ \rho\mu}+ \partial_\rho B_{\mu\nu} -\left( A_\mu F_{\nu \rho}+
A_\nu F_{ \rho\mu}+ A_\rho F_{\mu\nu}  \right)
\end{equation}
where $B_{\mu\nu}$  represents  the  stringy antisymmetric $B$-field. This action  has been considered as
the starting point to elaborate the Kerr-Sen (charged rotating) black hole in four dimensions.\\
In order  to investigate  the  corresponding thermodynamical  behaviors, however,  we exploit  its line element   metric
  using the standard Boyer-Lindquist coordinates. In this  way, it  reads as
\begin{equation}
ds^{2}=-\frac{f(r)}{\sigma}\left(dt-a sin^{2}\theta d\phi\right)^{2}+\frac{sin^{2}\theta}{\sigma}\left(a dt-(\sigma+a^{2}sin^{2}
\theta d\phi \right)^{2}+\sigma\left(\frac{dr^{2}}{f(r)}+d \theta^{2} \right).
\end{equation}
In this case, the metric function $f(r)$  reads as
  \begin{equation}
  f(r)= r^{2}+
  2(b-M)r+a^{2}.
  \end{equation}
The quantity $\sigma$ takes the following form
\begin{equation}
\sigma=r^{2}+2br+a cos^{2}\theta.
\end{equation}
The twist parameter $b$ and the angular momentum $J$ are respectively given by
\begin{equation}
\label{bJ}
b=\frac{Q^{2}}{2M},     \hspace{1.5cm} J=aM.
\end{equation}
In this  Kerr-Sen black hole solution,  $M$, $Q$ and $a$   represent the mass, the  charge and  the specific angular momentum,  respectively.
 As usually, the event horizon $r_h$ of such a black  hole can be obtained  by  imposing the constraint $f(r_h) = 0$. The associated solution  is given by
\begin{equation}
r^{\pm}_h=M-b\pm\sqrt{(M-b)^{2}-a^{2}}.
\end{equation}
Using Eq(\ref{bJ}),   the corresponding  entropy reads as
\begin{equation}
S=\frac{A}{4}=\pi\left( 2M^2-Q^{2}+\sqrt{\left(2M^2-Q^{2}\right)^{2}-4J^{2}}\right).
\end{equation}
The Hawking temperature,  being related to the  surface gravity,  takes the following form
\begin{equation}
T=\frac{\sqrt{\left(2M-Q^{2}\right)^{2}-4J^{2}}}{4\pi M\left( 2M-Q^{2}+\sqrt{\left(2M-Q^{2}\right)^{2}-4J^{2}}\right)}.
\end{equation}
Similar calculations provide a generalized formula for  the Kerr-Sen black hole
\begin{equation}
M(S,Q,J)=\left(\frac{4\pi^2 J^2 + 2 \pi Q^2 S + S^2}{4 \pi S}\right)^{1/2}.
\end{equation}
It is recalled that  the  first law of  the  black hole thermodynamics can be written as
\begin{equation}
dM=TdS + \phi dQ + \Omega_{H} dJ,
\end{equation}
where $T$, $\phi$ and $ \Omega_{H}$ are the temperature,  the electrostatic potential and  the angular velocity, respectively.
The explicit expressions of the involved  parameters are  listed as follows
\begin{eqnarray}
\Omega &=& \frac{\partial M}{\partial J}=\frac{\pi J}{M S}
\label{4}
\\
\phi&=&\frac{\partial M}{\partial Q}=\frac{Q}{2M}
\label{6}.
\end{eqnarray}
The  temperature being given by
\begin{eqnarray}
T(S,Q,J)&=&\frac{\partial M}{\partial S}=\frac{(S-2\pi J)(S+2\pi J)}{4S^{3/2} \sqrt{\pi(4 \pi^2 J^2 S + 2\pi Q^2 S^2 + S^3})}
\label{5}
\end{eqnarray}
 is illustrated in Fig.\ref{1} as a function of the entropy for different values of the angular momentum and the charge.
 This computation  generates certain critical behaviors.
\\
A close examination  on  the curve in the  $S-T$   plane   shows  that  the temperature  $T$ is indeed a continuous function
of the entropy. Thus,  the possibility of a first-order phase transition is removed. Besides, one can notice from Fig.\ref{1}
 that the temperature decreases when the angular momentum and the charge does.   However,
it has been shown  that    the temperature  vanishes at $S=2\pi J$  which could be linked with   an  extremal black hole solution. Then,  the temperature exhibits
 a maximum point corresponding to  its
vanishing first derivative  located at
\begingroup\makeatletter\def\f@size{10}\check@mathfonts
\begin{equation}
S_{max}=\pi  \sqrt{2 \sqrt[3]{2 J^4 Q^4-8 J^6}+4 J^2}+\sqrt{4 \pi ^2 J^2 \left(\frac{Q^2}{\sqrt{\frac{1}{2} \sqrt[3]{2 J^4 Q^4-8 J^6}+J^2}}+2\right)-2 \pi ^2
   \sqrt[3]{2 J^4 Q^4-8 J^6}}.
\end{equation}
\endgroup
Then,  it goes to  zero when  $S$  goes to infinity. It is  known that a second-order occurs at  the point where
 the heat capacity  exhibits  a singularity.
To get further insight into the thermodynamical behaviors of the  Kerr-Sen black hole,  we compute  the semi-classical specific
 heat by using the relation $ C_{J,Q}=T\left(\frac{\partial S}{\partial T}\right)$.
  This has been found to be
\begin{equation}
C_{J,Q}(S,Q,J)= \frac{2 S (-4 J^2 \pi^2 + S^2) (4 J^2 \pi^2 + S (2 \pi Q^2 + S))}{48 J^4 \pi^4 - S^4 + 8 J^2 \pi^2 S (4 \pi Q^2 + 3 S)}.
\label{21}
\end{equation}
To investigate  the associated  behaviors, we plot the heat  capacity as a function of the entropy  in Fig.\ref{2}.

It follows   that this function involves  a discontinuity  at a critical
 value of the  entropy  given by  $S_{c}=S_{max}=1.60199$.  Such a point  $S_{c}$ separates  two branches associated with a  small
  and large black hole transition.
It can be seen  also  from   Fig.\ref{2} that the heat capacity  changes from  a positive  infinity to  a negative infinity at $T = T_c=0.0874147$.
 An examination shows  that  there is an infinite divergence of the heat capacity that indicates a higher-order phase
 transition of the Kerr-Sen black hole. This critical transition
    can be treated using  two different approaches which will be  investigated in the next sections. The first one will be based on the   Ehrenfest method. However,
    the second  one will be conducted using a geometric  way by adopting a new metric form   inspired by the Gibbs free energy scaled  by a  conformal
    factor associated with the existence of  extremal solutions.

\section{Second-order phase transition of   the  Kerr-Sen  black hole}
In this section, we study   a  phase transition phenomena  appearing in the  Kerr-Sen black hole by  using first the
Ehrenfest scheme  \cite{w20,w21,w22,w23,w24}. In particular, we will   show that this black hole exhibits a second-order  phase
 transition by adopting such  a nice  technique
used for studying  thermodynamical  systems.
\subsection{Kerr-Sen  black hole behaviors from  the Ehrenfest scheme}
\begin{figure}[!ht]
		\begin{center}
		\centering
			\begin{tabbing}
			\centering
			\hspace{8.5cm}\=\kill
			\includegraphics[scale=.75]{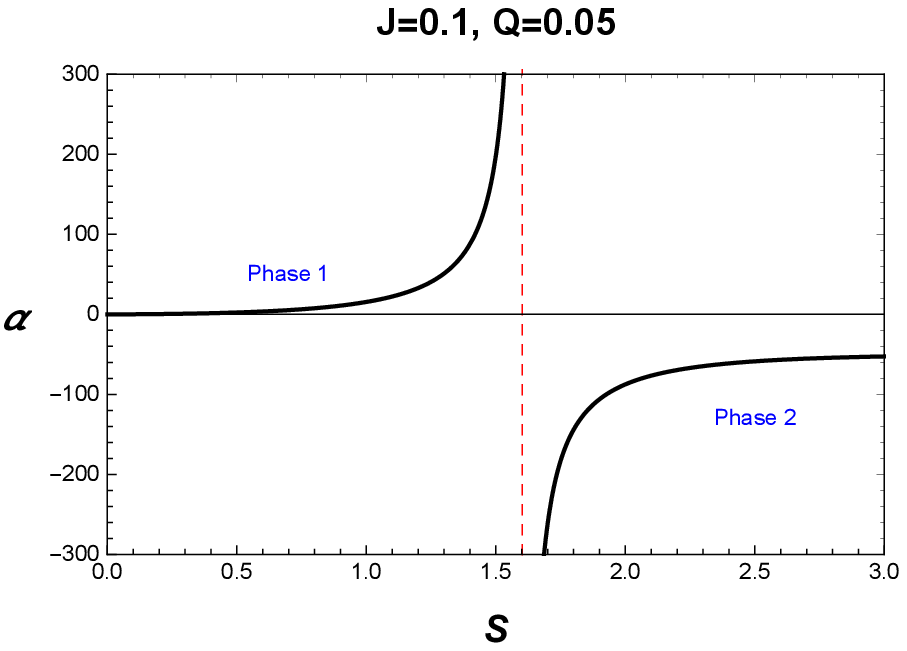} \>
			\includegraphics[scale=.75]{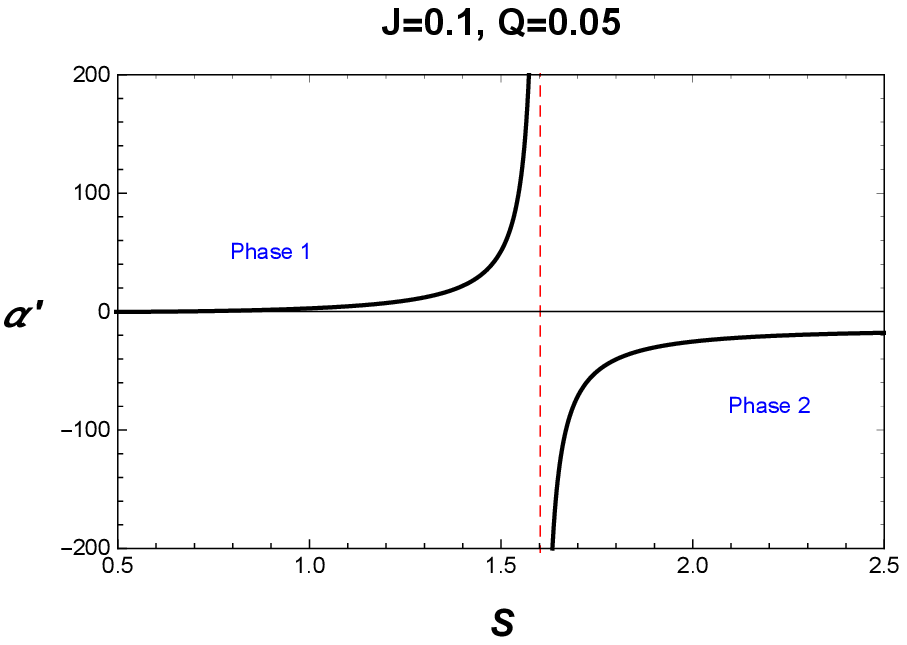} \\
			\includegraphics[scale=.75]{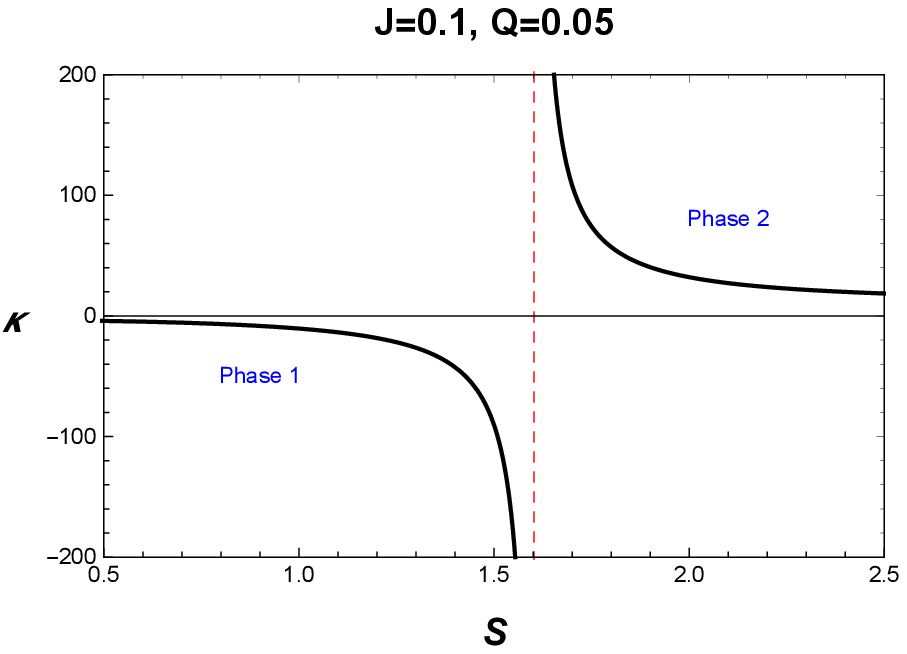} \>
			\includegraphics[scale=.75]{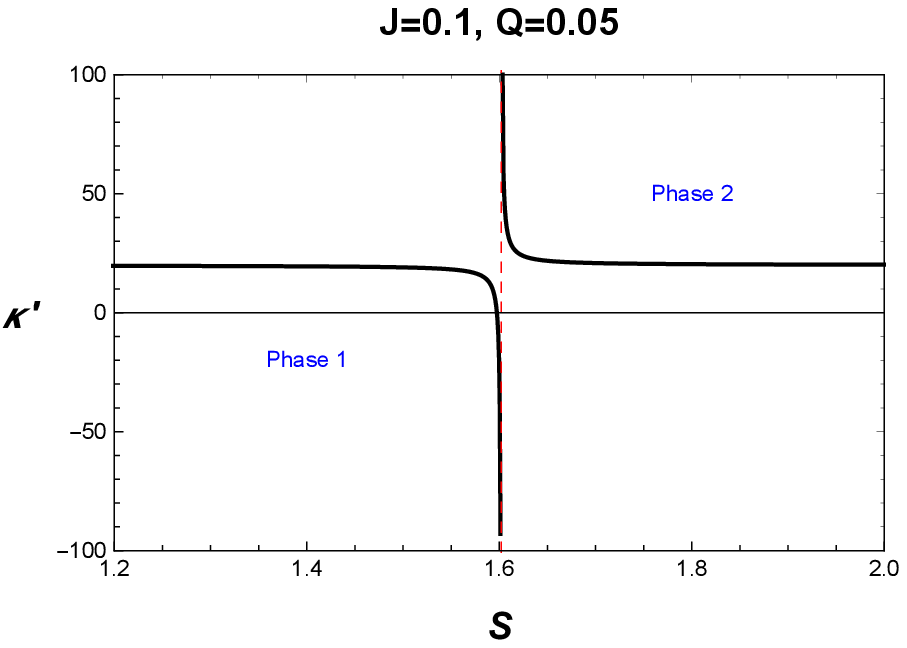} \\
			\includegraphics[scale=.75]{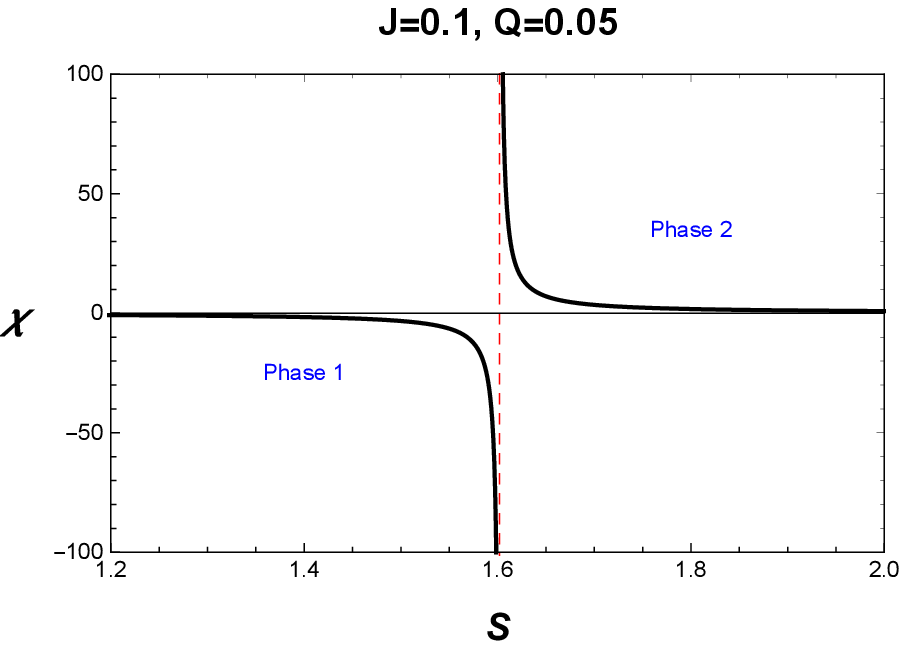} \>
			\includegraphics[scale=.75]{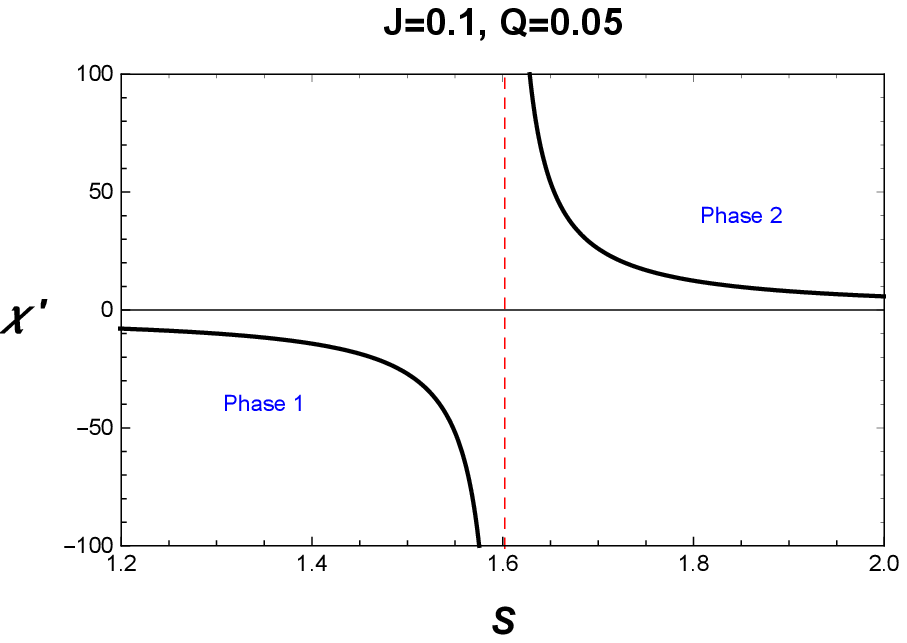} \\
		   \end{tabbing}
\caption{The variation of $\alpha$, $\kappa$,  $\chi$, $\alpha^{\prime}$, $\kappa^{\prime}$ and $\chi^{\prime}$ with respect to the entropy for fixed values of $Q$ and $J$.}
\label{13}
\end{center}
\end{figure}

It is noted that the  Ehrenfest scheme has been  exploited  to  understand the nature of the phase transition,  based on certain  relations and equations.
For such a black hole, twelve Ehrenfest equations are involved. They are  grouped in Tab.\ref{3}. \\
\begin{table}[h]
\centering
\begin{tabular}{|c|c|c|}
  \hline
  S fixed & $ \Omega$ fixed & $ \phi$ fixed \\
  \hline
$ -\left(\frac{\partial J}{\partial T}\right)_{S,Q} =\frac{({C_{J,Q}})_{2}-({C_{J,Q}})_{1}}{\Omega T(\alpha_{2}-\alpha_{1})} $ &
$ \left(\frac{\partial J}{\partial T}\right)_{\Omega,Q} =\frac{\alpha_{2}-\alpha_{1}}{\kappa_{2}-\kappa_{1}} $ &
 $ \left(\frac{\partial J}{\partial T}\right)_{\phi,Q} =\frac{\phi({\alpha^{\prime}}_{2}-{\alpha^{\prime}}_{1})}{\Omega(\chi_{2}-\chi_{1})} $    \\
  $ -\left(\frac{\partial J}{\partial Q}\right)_{S,T} =\frac{\phi({\alpha^{\prime}}_{2}-{\alpha^{\prime}}_{1})}{\Omega(\alpha_{2}-\alpha_{1})} $
   & $-\left(\frac{\partial J}{\partial Q}\right)_{\Omega,T} =\frac{\phi({\chi^{\prime}}_{2}-{\chi^{\prime}}_{1})}{\Omega(\kappa_{2}-\kappa_{1})} $
    & $-\left(\frac{\partial J}{\partial Q}\right)_{\phi,T} =\frac{\phi({\kappa^{\prime}}_{2}-{\kappa^{\prime}}_{1})}{\Omega(\chi_{2}-\chi_{1})} $ \\
 $ -\left(\frac{\partial Q}{\partial J}\right)_{S,T} =\frac{\Omega (\alpha_{2}-\alpha_{1})}{\phi ({\alpha^{\prime}}_{2}-{\alpha^{\prime}}_{1})} $  &
 $ -\left(\frac{\partial Q}{\partial J}\right)_{\Omega,T} =\frac{\Omega (\kappa_{2}-\kappa_{1})}{\phi ({\chi^{\prime}}_{2}-{\chi^{\prime}}_{1})} $
      &  $ -\left(\frac{\partial Q}{\partial J}\right)_{\phi,T} =\frac{\Omega (\chi_{2}-\chi_{1})}{\phi ({\kappa^{\prime}}_{2}-{\kappa^{\prime}}_{1})} $
        \\
  $ -\left(\frac{\partial Q}{\partial T}\right)_{S,J} =\frac{({C_{J,Q}})_{2}-({C_{J,Q}})_{1}}{\phi T({\alpha^{\prime}}_{2}-{\alpha^{\prime}}_{1})} $
   & $\left(\frac{\partial Q}{\partial T}\right)_{\Omega,J} =\frac{\Omega (\alpha_{2}-\alpha_{1})}{\phi ({\chi^{\prime}}_{2}-{\chi^{\prime}}_{1})} $&
   $ \left(\frac{\partial Q}{\partial T}\right)_{\phi,J} =\frac{{\alpha^{\prime}}_{2}-{\alpha^{\prime}}_{1}}{{\kappa^{\prime}}_{2}-{\kappa^{\prime}}_{1}} $
    \\
  \hline
\end{tabular}
\caption{Ehrenfest equations for the  Kerr-Sen black hole.}
\label{3}
\end{table}\\
The relevant thermodynamical coefficients are given by the following formulae
\begin{equation}
\begin{array}{ll}
\alpha=-\frac{1}{\Omega}{\left(\frac{\partial \Omega}{\partial T}\right)}_{J,Q},
 \hspace{1.2cm} \kappa=\frac{1}{\Omega}{\left(\frac{\partial \Omega}{\partial J}\right)}_{T,Q},
 \hspace{1.2cm} \chi=\frac{1}{\Omega}{\left(\frac{\partial \phi}{\partial J}\right)}_{T,Q}, \\ [6px]
\alpha^{\prime}=-\frac{1}{\phi}{\left(\frac{\partial \phi}{\partial T}\right)}_{J,Q},
 \hspace{1.2cm} \chi^{\prime}=\frac{1}{\Omega}{\left(\frac{\partial \Omega}{\partial Q}\right)}_{T,J}, \hspace{1.2cm}
  \kappa^{\prime}=\frac{1}{\Omega}{\left(\frac{\partial \phi}{\partial Q}\right)}_{T,J}.
\end{array}
\label{eqthco}
\end{equation}
As shown  in  Fig.\ref{2},   the heat capacity $C_{J,Q}$ is  discontinuous which is needed to reveal  that the studied black  hole system  undergoes
 a second-order phase transition.  This can be confirmed by examining the discontinuous behaviors of
   $\alpha$,  $\kappa$, $\chi$, $\alpha^{\prime}$, $\kappa^{\prime}$ and $\chi^{\prime}$.  To do so,
   one should  compute such quantities. Using \eqref{4} and \eqref{6}, we can express $\Omega$ and $\phi$ as
    a function of $S,J$ and $Q$.  The calculations  give
\begin{align}
& \Omega(S,Q,J)= \frac{2 J \pi^{3/2}}{\sqrt{2 \pi S^2 Q^2 + 4 J^2 s \pi^2 + S^3}}, \label{7} \\
& \phi(S,Q,J)= \frac{\sqrt{S\pi}Q}{\sqrt{4 J^2 \pi^2 + 2 \pi Q^2 S + S^2}}. \label{8}
\end{align}
By using \eqref{5}, \eqref{7}, and \eqref{8}, the expressions of $\alpha$ and $\alpha^\prime$ can be written as
\begin{align}
\label{15}
& \alpha=-\frac{1}{\Omega}{\left(\frac{\partial \Omega}{\partial S}\right)}_{J,Q}{\left(\frac{\partial S}{\partial T}\right)}_{J,Q}=\frac{4S (4 J^2 \pi^2 + S (4\pi Q^2 + 3 S))\sqrt{S \pi (4 J^2\pi^2 + S (2 \pi Q^2 + S))}}{48 J^4 \pi^4 - S^4 + 8 J^2 \pi^2 S (4 \pi Q^2 + 3 S)} \\
\label{16}
& \alpha^{\prime}=-\frac{1}{\phi}{\left(\frac{\partial \phi}{\partial S}\right)}_{J,Q}{\left(\frac{\partial S}{\partial T}\right)}_{J,Q}=\frac{4  S (-4 J^2 \pi^2 + S^2)\sqrt{S\pi(4 J^2 \pi^2 + S (2 \pi Q^2 + S))} }{48 J^4 \pi^4 - S^4 + 8 J^2 \pi^2 S (4 \pi Q^2 + 3 S)}.
\end{align}
%\begin{figure}
%\begin{center}
 %\includegraphics[scale=0.5]{alpha.pdf}
 %\caption{$\alpha$ with respect to the entropy (S) for fixed $J=0.1 $ and $Q=0.05$},
 %\label{9}
 %\end{center}
%\end{figure}
%\\
%\begin{figure}
%\begin{center}
 %\includegraphics[scale=0.5]{alphaprime.pdf}
 %\caption{$\alpha^{\prime}$ with respect to the entropy (S) for fixed $J=0.1 $ and $Q=0.05$},
 %\label{10}
 %\end{center}
%\end{figure}
Using the following equations
\begin{equation}
 {\left(\frac{\partial \Omega}{\partial J}\right)}_{T,Q}=\frac{{\left(\frac{\partial \Omega}{\partial J}\right)}_{S,Q}{\left(\frac{\partial T}{\partial S}\right)}_{J,Q}-{\left(\frac{\partial \Omega}{\partial S}\right)}_{J,Q}{\left(\frac{\partial T}{\partial J}\right)}_{S,Q}}{{\left(\frac{\partial T}{\partial S}\right)}_{J,Q}}, \;\;
 {\left(\frac{\partial \phi}{\partial Q}\right)}_{T,J}=\frac{{\left(\frac{\partial \phi}{\partial Q}\right)}_{S,J}{\left(\frac{\partial T}{\partial S}\right)}_{J,Q}-{\left(\frac{\partial \phi}{\partial S}\right)}_{J,Q}{\left(\frac{\partial T}{\partial Q}\right)}_{S,J}}{{\left(\frac{\partial T}{\partial S}\right)}_{J,Q}} \nonumber
\end{equation}
we obtain
 \begin{align}
\kappa=\frac{(4 J^2 \pi^2 + S^2)^{2}}{J(48 J^4 \pi^4 - S^4 + 8 J^2 \pi^2 S (4 \pi Q^2 + 3 S))}, \label{17} \\
\kappa^{\prime}=\frac{48 J^4\pi^4 + 24 J^2 \pi^2 S^2 - S^4}{Q(48 J^4 \pi^4 - S^4 + 8 J^2 \pi^2 S (4 \pi Q^2 + 3 S))}. \label{18}
\end{align}
To get the expressions of the two final quantities, we exploit the formulae
\begin{equation}
               {\left(\frac{\partial \phi}{\partial J}\right)}_{T,Q}=\frac{{\left(\frac{\partial \phi}{\partial J}\right)}_{S,Q}{\left(\frac{\partial T}{\partial S}\right)}_{J,Q}-{\left(\frac{\partial \phi}{\partial S}\right)}_{J,Q}{\left(\frac{\partial T}{\partial J}\right)}_{S,Q}}{{\left(\frac{\partial T}{\partial S}\right)}_{J,Q}}, \;\;
 {\left(\frac{\partial \Omega}{\partial Q}\right)}_{T,J}=\frac{{\left(\frac{\partial \Omega}{\partial Q}\right)}_{S,J}{\left(\frac{\partial T}{\partial S}\right)}_{J,Q}-{\left(\frac{\partial \Omega}{\partial S}\right)}_{J,Q}{\left(\frac{\partial T}{\partial Q}\right)}_{S,J}}{{\left(\frac{\partial T}{\partial S}\right)}_{J,Q}},\nonumber
            \end{equation}
which gives the following relations
\begin{align}
\chi=\frac{-4\pi Q S (4 J^2 \pi^2 + S^2)}{48 J^4 \pi^4 - S^4 + 8 J^2 \pi^2 S (4 \pi Q^2 + 3 S)}, \label{19} \\
\chi^{\prime}=\frac{-8 J \pi^2(4 J^2 \pi^2 + S^2)}{48 J^4 \pi^4 - S^4 + 8 J^2 \pi^2 S (4 \pi Q^2 + 3 S)}. \label{20}
\end{align}
It is worth noting  that $\alpha$, $\chi$, $\alpha^{\prime}$, $\kappa^{\prime}$ and $\chi^{\prime}$ share the same denominator
 being exactly  the one of $C_{J,Q}(S,Q,J)$.  Thus,  such quantities  diverge at the same critical value of the entropy. This critical
  behavior  can be clearly  seen  from Fig.\ref{13} showing the variation  of such quantities  in terms  of the entropy $S$.
 Precisely, it has been observed that all the computed quantities exhibit a discontinuity at the critical value
  $S_{c}=1.60199$ illustrated by  the  red dashed line.  The  obtained  results show
    a genuine second-order phase transition  for  the Kerr-Sen black hole.

\subsection{Validity  of the Ehrenfest equations}
Now we are in  position to  inspect   the Ehrenfest  equations  organized in the  table \ref{3} by
  checking  their validity. It is noted that for any  second-order phase transition, the Ehrenfest equations should
    be satisfied at the critical point.  Indeed, we consider  first the  Ehrenfest equations  associated  with
    the fixed entropy classes. Using \eqref{5},  the  left hand side of the first   equation can be computed. This has been  found to be
 \begin{equation}
  -\left(\frac{\partial J}{\partial T}\right)_{S,Q}=\frac{\left(\frac{S}{\pi}\left(4 J^2 \pi^{2} +S (2 \pi Q^2 +
   S\right)\right)^{3/2}}{J \left(4 J^2 \pi^2 + S (4 \pi Q^2 + 3 S)\right)}.
\label{23}
 \end{equation}
To   get  the right hand side, we  can use  the expressions of $C_{J,Q}$ and $ \alpha$ given  in  \eqref{21} and \eqref{15}.
 Indeed,    consider two points $S_{1}=S_{c}+\epsilon$ and $S_{2}=S_{c}-\epsilon$ $(\epsilon\ll 1)$ associated
  with phase 1 and phase 2, respectively of the curves given in Figs.\ref{2} and \ref{13} and the corresponding
  $C_{J,Q}\mid_{S_1} = {C_{(J,Q)_1}}$, $C_{J,Q}\mid_{S_2} = {C_{(J,Q)_2}}$
and $\alpha\mid_{S_1} =\alpha_{1}$, $\alpha\mid_{S_2} =\alpha_{2}$,  respectively.
Putting
 $D(S,J,Q)=48 J^4 \pi^4 - S^4 + 8 J^2 \pi^2 S (4 \pi Q^2 + 3 S)$, the right hand side of the first Ehrenfest  relation for fixed values of the  entropy reads as
\begin{equation}
\frac{({C_{J,Q}})_{2}-({C_{J,Q}})_{1}}{\Omega T(\alpha_{2}-\alpha_{1})}= \frac{2 S (-4 J^2 \pi^2 + S^2) (4 J^2 \pi^2 + S (2 \pi Q^2 + S))}
{\Omega T [4S (4 J^2 \pi^2 + S (4\pi Q^2 + 3 S))\sqrt{S \pi (4 J^2\pi^2 + S (2 \pi Q^2 + S))}]}\times\frac{\left(\frac{1}{D_{2}}
-\frac{1}{D_{1}}\right)}{\left(\frac{1}{D_{2}}-\frac{1}{D_{1}}\right) }
\end{equation}
 where $D_{1}$  and  $D_{2}$ denote  the denominators  associated with the  phase 1 and 2,  respectively. Using the expressions of $T$
   and $\Omega$, the right hand side of the first Ehrenfest  relation for  a fixed  value of the entropy is found to be
\begin{equation}
\frac{({C_{J,Q}})_{2}-({C_{J,Q}})_{1}}{\Omega T(\alpha_{2}-\alpha_{1})}=\frac{\left(\frac{S}{\pi}
\left(4 J^2 \pi^{2} +S (2 \pi Q^2 + S\right)\right)^{3/2}}{J \left(4 J^2 \pi^2 + S (4 \pi Q^2 + 3 S)\right)}
\label{r1}
\end{equation}
revealing  that the first Ehrenfest  equation is satisfied.  Using similar   calculations, we examine
 the second equation  appearing in  Tab.\ref{3}. The left hand side of the second Ehrenfest  equation can be written as
\begin{equation}
 -\left(\frac{\partial J}{\partial Q}\right)_{S,T} = \frac{{\left(\frac{\partial T}{\partial Q}
 \right)}_{S,J}}{{\left(\frac{\partial T}{\partial J}\right)}_{S,Q}}.
 \label{25}
\end{equation}
Using the equations \eqref{5} and \eqref{25}, we  get
\begin{equation}
 -\left(\frac{\partial J}{\partial Q}\right)_{S,T} = \frac{Q S (S^2 -4 J^2 \pi^2 )}{8 J^3 \pi^3 + 2 J \pi S (4\pi Q^2 + 3 S)}.
\end{equation}
Exploiting  the same procedure  used in the  verification of  the first equation right hand side, we obtain
\begin{equation}
\frac{\phi({\alpha^{\prime}}_{2}-{\alpha^{\prime}}_{1})}{\Omega(\alpha_{2}-\alpha_{1})}=\frac{\phi[4
 S (-4 J^2 \pi^2 + S^2)\sqrt{S\pi(4 J^2 \pi^2 + S (2 \pi Q^2 + S))}]}{\Omega[4S (4 J^2 \pi^2 + S (4\pi Q^2 + 3 S))\sqrt{S \pi
 (4 J^2\pi^2 + S (2 \pi Q^2 + S))}]}\times\frac{\left(\frac{1}{D_{2}}-\frac{1}{D_{1}}\right)}{\left(\frac{1}{D_{2}}-\frac{1}{D_{1}}\right) }.
\end{equation}
Using \eqref{8} and  \eqref{7},  we arrive to
\begin{equation}
\frac{\phi({\alpha^{\prime}}_{2}-{\alpha^{\prime}}_{1})}{\Omega(\alpha_{2}-\alpha_{1})}=\frac{Q S (S^2 -4 J^2 \pi^2 )}
{8 J^3 \pi^3 + 2 J \pi S (4\pi Q^2 + 3 S)}
\end{equation}
which reveals   clearly the validity of the  Ehrenfest second  equation at the critical point. For the third equation,
the left hand side can be calculated  using  \eqref{5} and \eqref{26}. Indeed, it is given by
\begin{equation}
  -\left(\frac{\partial Q}{\partial J}\right)_{S,T} = \frac{{\left(\frac{\partial T}{\partial J}\right)}_{S,Q}}
  {{\left(\frac{\partial T}{\partial Q}\right)}_{S,J}}= \frac{8 J^3 \pi^3 + 2 J \pi S (4\pi Q^2 + 3 S)}{Q S (S^2 -4 J^2 \pi^2 )}.
 \label{26}
\end{equation}
Similar computations provide
\begin{equation}
\frac{\Omega(\alpha_{2}-\alpha_{1})}{\phi({\alpha^{\prime}}_{2}-{\alpha^{\prime}}_{1})}=
\frac{\Omega [4S (4 J^2 \pi^2 + S (4\pi Q^2 + 3 S))\sqrt{S \pi (4 J^2\pi^2 +
 S (2 \pi Q^2 + S))}]}{\phi [4  S (-4 J^2 \pi^2 + S^2)\sqrt{S\pi(4 J^2 \pi^2 + S
 (2 \pi Q^2 + S))}]}\times\frac{\left(\frac{1}{D_{2}}-\frac{1}{D_{1}}\right)}{\left(\frac{1}{D_{2}}-\frac{1}{D_{1}}\right) }.
\end{equation}
Using the expression of $\phi$  from \eqref{8} and $\Omega$ from \eqref{7}, we  find
\begin{equation}
\frac{\Omega(\alpha_{2}-\alpha_{1})}{\phi({\alpha^{\prime}}_{2}-{\alpha^{\prime}}_{1})}=\frac{Q S (S^2 -4 J^2 \pi^2 )}{8 J^3 \pi^3 +
 2 J \pi S (4\pi Q^2 + 3 S)}.
\end{equation}
This indicates the validity of such an   Ehrenfest equation. For the last equation given in  Tab.\ref{3}
for a fixed entropy, we follow the same  calculations by finding first  the left hand side. Concretely,  it  has been  found to be
\begin{equation}
-\left(\frac{\partial Q}{\partial T}\right)_{S,J}=\frac{2\left(S\left(4 J^2 \pi^{2} +S (2 \pi Q^2 +
 S\right)\right)^{3/2}}{\sqrt{\pi}QS\left(S^{2}-4 J^2 \pi^2)\right)}.
\end{equation}
For the right hand side, the calculation shows that it involves  the same expression as the left handed side. Thus, we have
\begin{equation}
\frac{({C_{J,Q}})_{2}-({C_{J,Q}})_{1}}{\phi T({\alpha^{\prime}}_{2}-{\alpha^{\prime}}_{1})}=\frac{2\left(S\left(4 J^2 \pi^{2}
+S (2 \pi Q^2 + S\right)\right)^{3/2}}{\sqrt{\pi}QS\left(S^{2}-4 J^2 \pi^2)\right)} =  -\left(\frac{\partial Q}{\partial T}\right)_{S,J}.
\end{equation}
As expected, the  remaining equations,  given in Tab.\ref{3},  can be examined using  the same method.  Instead  of repeating
 the associated calculation,  we give  only the obtained  results.  For a  fixed angular momentum,    they are   presented as follows
\begin{align}
& \frac{\alpha_{2}-\alpha_{1}}{\kappa_{2}-\kappa_{1}}=\frac{4S J (4 J^2 \pi^2 + S (4\pi Q^2 + 3 S))\sqrt{S \pi (4 J^2\pi^2 + S (2 \pi Q^2 + S))}}{(4 J^2 \pi^2 + S^2)^{2}}={\left(\frac{\partial J}{\partial T}\right)}_{\Omega,Q}, \\
\label{r2}
& \frac{\phi({\chi^{\prime}}_{2}-{\chi^{\prime}}_{1})}{\Omega(\kappa_{2}-\kappa_{1})}=\frac{4 J \pi Q S}{4 J^2 \pi^2 + S^2}=-{\left(\frac{\partial J}{\partial Q}\right)}_{\Omega,T}, \\
& \frac{\Omega(\kappa_{2}-\kappa_{1})}{\phi({\chi^{\prime}}_{2}-{\chi^{\prime}}_{1})}=\frac{4 J^2 \pi^2 + S^2}{4 J \pi Q S}=-{\left(\frac{\partial Q}{\partial J}\right)}_{\Omega,T}  , \\
& \frac{\Omega (\alpha_{2}-\alpha_{1})}{\phi ({\chi^{\prime}}_{2}-{\chi^{\prime}}_{1})}= \frac{Q (4 J^2 \pi^2 + S^2)}{\sqrt{\frac{S}{\pi}(4 J^2 \pi^2 + S (2 \pi Q^2 + S))}}= {\left(\frac{\partial Q}{\partial T}\right)}_{\Omega,J}.
\end{align}
For a fixed electric potential, the Ehrenfest equations are also verified
\begin{align}
& \frac{\phi({\alpha^{\prime}}_{2}-{\alpha^{\prime}}_{1})}{\Omega(\chi_{2}-\chi_{1})}=\frac{S (4 J^2 \pi^2 - S^2)\sqrt{S (4 J^2 \pi^2 + S (2 \pi Q^2 + S))}}{2 J \pi^{3/2} (4 J^2 \pi^2 + S^2)}=  {\left(\frac{\partial J}{\partial T}\right)}_{\phi,Q}, \\
& \frac{\phi({\kappa^{\prime}}_{2}-{\kappa^{\prime}}_{1})}{\Omega(\chi_{2}-\chi_{1})}=-\frac{48 J^4 \pi^4 + 24 J^2 \pi^2 S^2 - S^4}{32 J^3 \pi^4 Q + 8 J \pi^2 Q S^2}=-{\left(\frac{\partial J}{\partial Q}\right)}_{\phi,T}, \\
& \frac{\Omega(\chi_{2}-\chi_{1})}{\phi({\kappa^{\prime}}_{2}-{\kappa^{\prime}}_{1})}=-\frac{32 J^3 \pi^4 Q + 8 J \pi^2 Q S^2}{48 J^4 \pi^4 + 24 J^2 \pi^2 S^2 - S^4}=-{\left(\frac{\partial Q}{\partial J}\right)}_{\phi,T}, \\
& \frac{{\alpha^{\prime}}_{2}-{\alpha^{\prime}}_{1}}{{\kappa^{\prime}}_{2}-{\kappa^{\prime}}_{1}}=\frac{4 Q S (-4 J^2 \pi^2 + S^2)\sqrt{S\pi(4 J^2 \pi^2 + S (2 \pi Q^2 + S))}}{48 J^4\pi^4 + 24 J^2 \pi^2 S^2 - S^4}={\left(\frac{\partial Q}{\partial T}\right)}_{\phi,J}.
\end{align}
It has been shown that Prigogine- Defay (PD) ratio $\Pi$  can be  considered as a tool
 to measure the deviation from the second
Ehrenfest equation \cite{w240}. Performing numerical calculations from   eqs.\eqref{r1} and \eqref{r2},  it  has been  found to be
\begin{equation}
\Pi= \frac{\Delta C_{J,Q} \Delta \kappa}{T_c \Omega_c (\Delta \alpha)^2}=1.
\end{equation}
Since all the Ehrenfest equations presented in Tab.\ref{3} are verified,
we can now confirm the existence of a second-order phase transition for  the Kerr-Sen black hole. Note that this matches perfectly with the second-order equilibrium transition discussed in \cite{w24,ehr2,ehr3}.

  To  consolidate this result,  we shall explore  the geothermodynamics by proposing a new metric  form  derived from the Gibbs free energy scaled
   by a conformal factor associated  with  the existence  of  extremal solutions.
\section{Kerr-Sen black hole geothermodynamics}
\begin{figure}[H]
\begin{center}
 \includegraphics[scale=0.7]{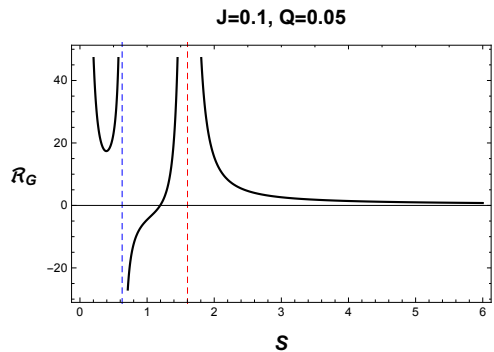}
 \caption{Scalar curvature as a function of the entropy (S).  The red dashed line represents the critical point at which   $\mathfrak{R}_G$ goes to infinity.  While,  the blue one is associated with the extremal black hole solution.}
 \label{X}
 \end{center}
\end{figure}

In this section, we   reconsider the investigation of    the  geothermodynamics, of such a black hole,   relying on
  singular  behaviors  of certain thermodynamical quantities including the heat capacity. The
   latter  is   relevant in  the determination of  the nature of  the  black hole  phase transitions,
    since  it involves  various  interesting thermodynamical properties. To unveil  such properties,
     the thermodynamical geometry of  the Kerr-Sen  black hole   has been approached by focusing on   the thermodynamical  geometric curvature.
     Several metrics could  be used  to  approach   such a quantity. However, here, we can explore an alternative  road.
  Inspired by  the  Hessian matrix of several free energies generated  from the Legendre transformation of $M$, we adopt a new metric form  derived from
  the  Gibss free energy  \cite{w25}. Concretely,  we introduce   a  conformal factor $\frac{1}{T}$ motivated by  extremal solutions being
   absent in known approaches. Exploiting the Gibbs free energy $G$,   we can write
\begin{align}
dG=-SdT+ \phi dQ + \Omega dJ.
\end{align}
Implementing a  conformal factor $\frac{1}{T}$, we obtain
\begin{equation}
\frac{1}{T}dG=\frac{1}{T} \left(-SdT + \phi dQ + \Omega dJ  \right).
\end{equation}
It  follows that the conformal Gibbs free energy metric can be written as
\begin{equation}
ds^2_G=\frac{1}{T} \left( -dT dS+ d\phi dQ + d\Omega dJ \right).
\end{equation}
Within the natural variables $T$, $Q$ and  $J$, it is convenient to  trade the temperature by the
entropy $S$. The resulting new  representation of the  metric takes the following form
\begin{equation}
ds^2_G=g_{SS}dS^2+g_{QQ}dQ^2+g_{JJ}dJ^2+2g_{QJ}dQdJ.
\end{equation}
With the use of this metric, we  try to check  the second-order phase transition point.
Thus, the scalar curvature can be expressed in the following way
\begin{equation}
\mathfrak{R}_G=\frac{\mathfrak{N}_G}{\mathfrak{D}_G},
\end{equation}
where the denominator is  found to be
\begin{equation}
\begin{aligned}
& \mathfrak{D}_G =  S(2J \pi -S)(2J \pi + S)(2\pi Q^2+S)^2(4J^2 \pi^2+S^2)(4J^2 \pi^2 + 2 \pi Q^2 S + S^2),  \\
&  (48 J^4 \pi^4 - S^4 + 8 J^2 \pi^2 S (4 \pi Q^2 + 3 S))^2,
\end{aligned}
\end{equation}
while the term $\mathfrak{N}_G$,  having a complicated  form, is   omitted since it is  not revelent in the present discussion. It is worth noting that
 the proposed metric gives a rather simple form of the scalar curvature.
Besides, the resulting scalar curvature has the same denominator as the quantities \eqref{21} and \eqref{eqthco}.
 This shows that the latter may also exhibit a discontinuity at the critical point $S_c=1.60199$  confirming the existence of  a
 second-order phase transition. To visualise such a result, we illustrate the scalar curvature in Fig.\ref{X} as a function of the entropy.

It follows from   this figure that the scalar is indeed discontinuous at $S_c$ (the red dashed line) and at $S=2 \pi J$ (the blue dashed line). This  not only
supports the result obtained in the previous section but also pushes one to investigate the associated scalings laws and critical  exponents.
\section{Scaling laws and critical  exponents}
In this section, we investigate  the scaling laws  and  the critical  exponents at the obtained  critical point  for   the Kerr-Sen black  hole. In the  elaboration of critical phenomena,  one can  meet with  three kinds of
 susceptibilities related to the heat capacity $C_A$, the  moment of inertia $(I_B)$ and  the electric capacitance $(K_c)$ defined as
\begin{equation}
C_A=-\beta^2 \left( \frac{\partial M}{\partial \beta} \right)_{A}, \quad I_B=\beta \left( \frac{\partial J}{\partial \left(\beta
\Omega \right)} \right)_{B}, \quad K_C=\beta \left( \frac{\partial Q}{\partial \left(\beta \phi\right)} \right)_{C}.
\label{CIK}
\end{equation}
Here, $\beta$ is the inverse of the  temperature and the set $\left\lbrace A, B, C \right\rbrace$ denotes the fixed quantities \cite{w26}. These quantities representing different exchange modes (i.e thermal, electrical and mechanical) show a behavioral  divergence  providing   a phase transition.
 It is noted that other  propositions have been conducted  for the phase transitions at the extremal limit,  being  a point of a second-order transition,
  when $M=a+b$ based on the divergence
  of thermal fluctuations.  Precisely, it has been remarked  that these fluctuations   can be linked to the divergence
   of \eqref{CIK}. Before going ahead, it is convenient to re-write  certain   needed thermodynamical quantities. In particular,  they
    are listed as follows
\begin{align}
& S_{\pm}=\frac{A_{\pm}}{4}= \pi \left[ 2M^2-Q^2 \pm \sqrt{\left( 2M^2-Q^2 \right)^2-4J^2} \right], \\
& T_{\pm}=\frac{\pm \sqrt{\left( 2M^2-Q^2 \right)^2-4J^2}}{M \times A_{\pm}} =\beta^{-1}, \\
& \Omega_{\pm}=\frac{4 \pi J}{M A_{\pm}}, \quad \phi_{\pm}= \frac{4 \pi Q r_h}{A_{\pm}}.
\end{align}
It is noted  that the entropy $S(M, J,Q)$ is a generalized homogeneous function since one has  \\$S(\sqrt{\lambda}M,
 \lambda J,\sqrt{\lambda}Q)=\lambda S(M, J,Q)$.
 The response coefficients are related to the susceptibilities via  the  following relations
\begin{align}
\label{chi1}
& \overline{\chi}_1 \equiv \left( \frac{\partial^2 S}{\partial M^2} \right)_{JQ}=-\frac{\beta^2}{C_{JQ}}, \\
\label{chi2}
& \overline{\chi}_2 \equiv \left( \frac{\partial^2 S}{\partial J^2} \right)_{MQ}=-\frac{\beta}{I_{MQ}}, \\
\label{chi3}
& \overline{\chi}_3 \equiv \left( \frac{\partial^2 S}{\partial Q^2} \right)_{MJ}=-\frac{\beta}{K_{MJ}}.
\end{align}
Near the critical points,  these quantities obey certain power laws. To derive such  laws, we  should define
 the following order parameters $\eta_M= \beta_+ - \beta_-$, $\eta_J= (\beta \Omega)_+ - (\beta \Omega)_-$ and
  $\eta_Q= (\beta \phi)_+ - (\beta \phi)_-$. It is worth noting that these quantities do not go to zero at the
   critical point. However, they  diverge since the inverse of the black hole temperature does. In this regime, we also need
   to define the following terms $M = M_X(1+\epsilon_M), J = J_X(1-\epsilon_J ) ,Q = Q_X(1-\epsilon_Q$), where
   $M_X, J_X$ and $Q_X$ represent the mass, the angular momentum, and the charge at the extremal limit,
     respectively and $\epsilon$ is associated with  small parameter deviations\footnote{
    In the extremal limit, we obtain $M_X=\sqrt{J+Q^2/2}, Q_X=\sqrt{2(J+M^2)}$ and $J_X=\frac{Q^2-2M^2}{2}$.}. Some of the relevant scaling laws are given by
\begin{align}
& \overline{\chi}_1 \sim \epsilon_M^{-\alpha} (J=0 \,\text{or} \, Q=0), \\
& \overline{\chi}_2 \sim \epsilon_M^{-\gamma} ( Q=0), \\
& \eta_J \sim \epsilon_M^{\beta} ( Q=0), \\
& \eta_J \sim \epsilon_M^{\delta-1} ( Q \neq 0).
\end{align}
Using the definition of $\overline{\chi}_1$ in the extremal limit where $M =\sqrt{J+Q^2/2}(1+\epsilon_M)$ and taking $Q=0$, we find
\begingroup\makeatletter\def\f@size{9}\check@mathfonts
\begin{align}
\small & \overline{\chi}_1 \approx \frac{ 4 \pi }{\epsilon_M ^{3/2} (\epsilon_M +2)^{3/2} (\epsilon_M  (\epsilon_M +2)+2)^{3/2}}  \\ \nonumber
& \times \left[ -2+\epsilon_M ^{3/2} (\epsilon_M +2) \left((\epsilon_M +2) \sqrt{\epsilon_M }  \left(\epsilon_M ^2+2 \epsilon_M +\sqrt{\epsilon_M } \sqrt{\epsilon_M +2} \sqrt{\epsilon_M  (\epsilon_M +2)+2}+3\right)+2 \sqrt{\epsilon_M +2}
  \sqrt{\epsilon_M  (\epsilon_M +2)+2}\right)\right].
\end{align}
\endgroup
Thus, we obtain  $\overline{\chi}_1 \propto \epsilon_M^{-3/2}.$  Using  similar  road calculations, we get
\begin{equation}
 \overline{\chi}_2 \approx -\frac{2 \pi  (\epsilon_M +1)^4}{J \left(\epsilon_M ^4+4 \epsilon_M ^3+6 \epsilon_M ^2+4 \epsilon_M \right)^{3/2}},
\end{equation}
which  shows that  $\overline{\chi}_2 \propto \epsilon_M^{-3/2}$. For the third equation, we obtain
\begin{equation}
\eta_J \approx -\frac{4 \pi }{\sqrt{\epsilon_M ^4+4 \epsilon_M ^3+6 \epsilon_M ^2+4 \epsilon_M }},
\end{equation}
revealing that  one has  $\eta_J  \propto \epsilon_M^{-1/2}$. Identical computations are  performed  for the last equation. The critical exponents for
 the Kerr-Sen black hole are therefore summarized as follows
\begin{equation}
\alpha=\frac{3}{2}, \, \, \, \gamma=\frac{3}{2}, \, \, \, \beta= -\frac{1}{2},  \, \, \, \delta=-2.
\end{equation}
From these critical exponents,  we  find  the equalities related to the scaling laws of the first kind
\begin{equation}
\alpha + 2 \beta + \gamma=2, \quad \beta \left( \delta-1 \right)=\gamma.
\end{equation}
This   matches perfectly with  the result of \cite{w26}  where   the  same relations  have been obtained for  the  Kerr-Newman black hole.
\section{Conclusions}
In this work, we have  investigated the phase transitions of the Kerr-Sen black hole engineered   from  the heterotic superstring theory.
In particular, we have computed the involved thermodynamical quantities.  It has been shown that the  phase transition
is characterized by divergences in the specific  heat and other quantities   near the critical points.  Using the  Ehrenfest  scheme,
 we have inspected  the nature of  the phase
transitions. Concretely,  we have analytically verified  the validity of  such equations  near
 the critical point. We have shown that  the  phase transition  corresponding  to the divergence in the heat capacity at
the critical point is  in fact   a second-order equilibrium  one. To support such a finding, we have approached  the
   Kerr-Sen black hole geothermodynamics  by adopting a new metric form derived from  the Gibbs  free energy scaled  by
    a conformal  factor  inspired by  the  singularity associated with  extremal solutions. This
   geometric method  recovers  similar behaviors. At the end,  we  have  obtained  the  involved  scaling
      behaviors  at the critical point.  \\ This work comes up with some  open questions.
      It  will be interesting to extend  this investigation by   considering       backgrounds with  a non  zero  cosmological  constant   built  recently in  \cite{w27}.
       Motivated by the black hole observational aspect such the gravitational waves
        and black hole image, a  second   investigation  could concern   optical properties
        from   different backgrounds, including dark sectors.
We leave these questions for future works.

\section*{Acknowledgment}
 This work is partially supported by the ICTP through AF-13.


\begin{thebibliography}{24}

\bibitem{w1} S. W. Hawking, \textit{Black holes in general relativity}, Communications in Mathematical Physics \textbf{25}(2) (1972) 152-166.

\bibitem{w2} J. B. Hartle, T. Dray, \textit{Gravity: An introduction to Einsteins general relativity}, Amer. J. Phys. \textbf{71} (2003) 1086-1087.

\bibitem{w3} J. L. Zhang, R. G. Cai, H. Yu, \textit{Phase transition and thermodynamical geometry for
Schwarzschild AdS black hole in $AdS_5 \times S^5$ spacetime}, J. High Ener. Phys. \textbf{2} (2015) 143.

\bibitem{w4} A. Belhaj, M. Chabab, H. El Moumni, K. Masmar, M. B. Sedra, \textit{On thermodynamics of AdS black holes in M-theory}, Eur. Phys. J. C \textbf{76}(2) (2016) 73.

\bibitem{w5} A. Belhaj, A. El Balali, W. El Hadri, Y. Hassouni, E. Torrente-Lujan, \textit{Phase
Transitions of Quintessential AdS Black Holes in M-theory/Superstring Inspired Models}, arXiv:2004.10647.


\bibitem{w6} H. J. Boonstra, B. Peeters, K. Skenderis, \textit{Branes And Anti-de Sitter Space-times}, Prog. Phys. \textbf{47}(1-3) (1999) 109-116.

\bibitem{w7} A. Sen, \textit{Rotating charged black hole solution in heterotic string theory}, Phys. Rev. Lett.\textbf{69}(7) (1992) 1006.


\bibitem{w8} S. Liebes Jr, \textit{Gravitational lenses}, Phys. Rev. \textbf{B835} (1964)133 .

\bibitem{w9} L. Ming-Jian, \textit{Temperature and thermodynamic geometry of the Kerr-Sen black hole}, Chin. Phys. B. \textbf{20}.2 (2011) 020404.

\bibitem{w10} X. G. Lan, J. Pu, \textit{Observing the contour profile of a Kerr-Sen black hole}, Mod. Phys. Lett. A. \textbf{33}.17 (2018) 1850099.

\bibitem{ww}  A. Belhaj, M. Benali, A. El Balali, H. El Moumni, S-E. Ennadifi, \textit{Deflection Angle and Shadow Behaviors of Quintessential Black
 Holes in arbitrary Dimensions}, arXiv:2006.01078.

\bibitem{w11} K. Akiyama et al. [Event Horizon Telescope Collaboration], \textit{First M87 Event Horizon
Telescope Results. I. The Shadow of the Supermassive Black Hole}, Astrophys. J. \textbf{875} (2019) no.1, L1.



\bibitem{w12} A. Belhaj, M. Chabab, H. El Moumni and M. B. Sedra, \textit{On thermodynamics of AdS black holes in arbitrary dimensions}, Chin. Phys. Lett. \textbf{29} (2012) 100401.

\bibitem{w13} A. Belhaj, M. Chabab, H. El Moumni, L. Medari and M. B. Sedra, \textit{The thermodynamical
behaviors of KerrNewman AdS black holes}, Chin. Phys. Lett. \textbf{30} (2013) 090402.


\bibitem{w14} S. W. Hawking, \textit{Black holes and thermodynamics}, Phys. Rev. D \textbf{13}(2) (1976) 191.

\bibitem{w15} J. Xu, L. M. Cao, Y. P. Hu, \textit{P-V criticality in the extended phase space of black holes
in massive gravity}, Phys. Rev. D \textbf{91}(12) (2015) 124033.

\bibitem{w16} A. Belhaj, A. El Balali,  W. El Hadri, M. A. Essebani, M. B.Sedra ,
 A. Segui, \textit{Kerr-AdS Black Hole Behaviors from Dark Energy}, International Journal of Modern Physics D, doi:
10.1142/S0218271820500698.

\bibitem{w17} Y. Wu and W. Xu, \textit{Effet of dark energy on Hawking-Page transition}, Phys.
Dark Univ. \textbf{27} (2020) 100470.

\bibitem{ww2} A. Belhaj, A. El Balali, W. El Hadri, H. El Moumni, M. B. Sedra, \textit{Dark
energy effects on charged and rotating black holes}, Eur. Phys. J. Plus \textbf{134}(9) (2019) 422.

\bibitem{w18} J.X. Mo, W. B. Liu. \textit{Ehrenfest scheme for PV criticality in the extended phase space of black holes}, Phys. Lett.
 B \textbf{727}.1-3 (2013) 336-339.

\bibitem{w19} A. Lala,  D. Roychowdhury, \textit{Ehrenfest’s scheme and thermodynamic geometry in
Born-Infeld AdS black holes}, Phys. Rev. D \textbf{86} 8 (2012) 084027.


\bibitem{w20} H. E. Stanley, \textit{Introduction to phase transitions and critical phenomena}, Oxford University Press, New York (1987).

\bibitem{w21} M. W. Zemansky and R. H. Dittman, \textit{Heat and thermodynamics: an intermediate}
textbook, McGraw-Hill (1997).

\bibitem{w22} R. Banerjee, S. K. Modak and S. Samanta, \textit{Glassy phase transition and stability in black holes}, Eur. Phys. J. C {\bf 70}  (2010)317.

\bibitem{w23} R. Banerjee, S. Ghosh and D. Roychowdhury, \textit{New type of phase transition in Reissner Nordström–AdS black hole and its thermodynamic geometry}, Phys. Lett.B {\bf 696} (2011) 156.

\bibitem{w24} T. M. Nieuwenhuizen, \textit{Ehrenfest relations at the glass transition:  solution to anold paradox}, Phys. Rev. Lett. {\bf 79}(1997)1317-1320.


\bibitem{w240} R. Banerjee, D. Roychowdhury, \textit{Thermodynamics of phase transition in higher dimensional AdS black holes}, JHEP{\bf 11}(2011)004.

%\bibitem{ehr1} T. M. Nieuwenhuizen, \textit{Ehrenfest relations at the glass transition:  solution to anold paradox}, Phys. Rev. Lett {\bf79}(1997)7.


%\cite{Xu:2014tja}
\bibitem{ehr2}
H.~Xu, W.~Xu and L.~Zhao, \textit{Extended phase space thermodynamics for third order Lovelock black holes in diverse dimensions}, Eur. Phys. J. C \textbf{74}, no.9, (2014)3074,
%doi:10.1140/epjc/s10052-014-3074-1
arXiv:1405.4143.
%77 citations counted in INSPIRE as of 13 Jul 2020


%\cite{Belhaj:2014tga}
\bibitem{ehr3}
A.~Belhaj, M.~Chabab, H.~EL Moumni, K.~Masmar and M.~B.~Sedra, \textit{Ehrenfest scheme of higher dimensional AdS black holes in the third-order Lovelock–Born–Infeld gravity}, Int. J. Geom. Meth. Mod. Phys. \textbf{12}, no.10, (2015)1550115,
%doi:10.1142/S0219887815501157
arXiv:1405.3306.
%32 citations counted in INSPIRE as of 13 Jul 2020

\bibitem{w25} H. Liu, H. Lu, M. Luo, K.N. Shao, \textit{Thermodynamical metrics and black hole phase
transitions}, JHEP {\bf 1012}  (2010) 054.

\bibitem{w26}  O. Kaburaki, \textit{Scaling laws at the critical point on black hole equilibrium series},  Phys.  Lett.  A {\bf 217} (1996)315-320.

\bibitem{w27}  D. Wu, P. Wu, H. Yu, S-Q.  Wu, \textit{Are ultra-spinning Kerr-Sen-AdS4 black holes always super-entropic?}, {\tt arXiv:2007.02224}.




\end{thebibliography}
\end{document}